\documentclass[12pt]{article}
\usepackage{amsmath,amssymb,amsthm,mathrsfs,url,hyperref}
\usepackage[usenames,dvipsnames]{xcolor}
\usepackage[normalem]{ulem}
\usepackage{cancel}

\title{Existence of Schr\"odinger Evolution with Absorbing Boundary Condition}
\author{
Lawrence Frolov\footnote{Department of Mathematics, Rutgers University, 110 Frelinghuysen Road, Piscataway, NJ 08854-8019, USA}~\footnote{E-mail: laf230@math.rutgers.edu},~~Stefan Teufel\footnote{Mathematisches Institut, 
	Eberhard-Karls-Universit\"at, 
	Auf der Morgenstelle 10, 72076 T\"ubingen, 
	Germany.}~\footnote{E-mail: stefan.teufel@uni-tuebingen.de},~~and  
Roderich Tumulka${}^\ddagger$\footnote{E-mail: roderich.tumulka@uni-tuebingen.de}
}
\date{July 3, 2025}

\newcommand{\Hilbert}{\mathscr{H}}
\newcommand{\be}{\begin{equation}}
\newcommand{\ee}{\end{equation}}
\newcommand{\scp}[2]{\langle #1|#2\rangle}
\renewcommand{\Re}{\mathrm{Re}}
\renewcommand{\Im}{\mathrm{Im}}
\newcommand{\CCC}{\mathbb{C}}

\newcommand{\NNN}{\mathbb{N}}
\newcommand{\RRR}{\mathbb{R}}
\newcommand{\SSS}{\mathbb{S}}

\newcommand{\vj}{{\boldsymbol{j}}}
\newcommand{\vn}{\boldsymbol{n}}

\newcommand{\vx}{\boldsymbol{x}}

\newcommand{\vX}{{\boldsymbol{X}}}

\newcommand{\R}{\Omega}
\newcommand{\bou}{\partial \R}
\newcommand{\prob}{\mathrm{Prob}}
\newcommand{\sA}{\mathscr{A}}
\newcommand{\sB}{\mathscr{B}}

\newcommand{\sI}{\mathscr{I}}
\newcommand{\sJ}{\mathscr{J}}
\newcommand{\sZ}{\mathscr{Z}}
\newcommand{\RT}[1]{{#1}}
\newcommand{\LFn}{}

\newcommand{\Hi}{_{L^2(\Omega)}}

\theoremstyle{plain}
\newtheorem{thm}{Theorem}
\newtheorem{prop}{Proposition}
\newtheorem{lm}{Lemma}
\newtheorem{cor}{Corollary}

\theoremstyle{definition}
\newtheorem{rem}{Remark}
\newtheorem*{definition}{Definition}
\begin{document}
\maketitle
\begin{abstract}
Consider a non-relativistic quantum particle with wave function inside a region $\R\subset \RRR^3$, and suppose that detectors are placed along the boundary $\partial \R$. The question how to compute the probability distribution of the time at which the detector surface registers the particle boils down to finding a reasonable mathematical definition of an ideal detecting surface; a particularly convincing definition, called the \emph{absorbing boundary rule}, involves a time evolution for the particle's wave function $\psi$ expressed by a Schr\"odinger equation in $\R$ together with an ``absorbing'' boundary condition on $\partial \R$ first considered by Werner in 1987, viz., $\partial \psi/\partial n=i\kappa\psi$ with $\kappa>0$ and $\partial/\partial n$ the normal derivative. We provide here a discussion of the rigorous mathematical foundation of this rule. First, for the viability of the rule it plays a crucial role that these two equations together uniquely define the time evolution of $\psi$; we point out here how, under some technical assumptions on the regularity (i.e., smoothness) of the detecting surface, the  Lumer-Phillips theorem implies that the time evolution is well defined and given by a contraction semigroup. Second, we show that the collapse required for the $N$-particle version of the problem is well defined. We also prove that the joint distribution of the detection times and places, according to the absorbing boundary rule, is governed by a {\LFn  positive-operator-valued measure}.

\medskip

\end{abstract}

\section{Introduction}

Suppose an ideal detecting surface is placed along the boundary $\bou$ of an open region $\R\subset \RRR^3$ in physical space, and a non-relativistic quantum particle is prepared at time 0 with wave function $\psi_0$ with support in $\R$. Let $Z=(T,\vX)\in [0,\infty)\times \bou$ be the random time and location of the detection event; we write $Z=\infty$ if no detection event ever occurs. What is the probability distribution of $Z$? As we have argued elsewhere \cite{detect-rule}, there is a simple rule for computing this distribution that is particularly convincing, called the absorbing boundary rule; its equations were first considered by Werner \cite{Wer87}. According to this rule, $\psi$ evolves according to the Schr\"odinger equation
\be\label{Schr}
i\hbar\frac{\partial \psi}{\partial t} = -\frac{\hbar^2}{2m} \nabla^2 \psi + V\psi
\ee
in $\R$ with potential $V:\R\to\RRR$ and boundary condition
\be\label{bc}
\frac{\partial \psi}{\partial n}(\vx) = i\,\kappa(\vx) \psi(\vx)
\ee
at every $\vx\in\bou$, where $\partial/\partial n$ is the outward normal derivative on the surface, i.e.,
\be
\frac{\partial \psi}{\partial n}(\vx) := \vn(\vx) \cdot \nabla\psi(\vx)
\ee
with $\vn(\vx)$ the unit vector perpendicular to $\bou$ at $\vx\in\bou$ pointing outside $\R$, and $\kappa(\vx)\geq0$ are given values of dimension 1/length that characterize the type of ideal detector (wave number of sensitivity). 

Then, the absorbing boundary rule asserts,
\begin{equation}\label{probnjR}
  \prob_{\psi_0} \Bigl( t_1 \leq T<t_2, \vX \in B \Bigr) =
  \int\limits_{t_1}^{t_2} dt \int\limits_{B} d^2\vx \; \vn(\vx) \cdot
  \vj^{\psi_t}(\vx)
\end{equation}
for any $0\leq t_1<t_2$ and any measurable set $B\subseteq \bou$, with $d^2\vx$ the surface area element and $\vj^\psi$ the probability current vector field
defined by $\psi$, which is
\begin{equation}\label{jSchr}
  \vj^\psi = \frac{\hbar}{m} \Im\, \psi^* \nabla \psi\,{\LFn ,}
\end{equation}
{\LFn where $*$ means complex conjugation.} Note that the boundary condition \eqref{bc} implies that the current $\vj^{\psi}$ is always outward-pointing on $\bou$, i.e., $\vn(\vx)\cdot \vj^{\psi}(\vx) \geq 0$, so \eqref{bc} is an ``absorbing'' boundary condition, and one should expect $\|\psi_t\|$ not to be constant but to be a decreasing function of $t$. It is taken for granted in \eqref{probnjR} that $\|\psi_0\|=1$. Finally, to complete the statement of the absorbing boundary rule, the probability that no detection ever occurs is
\be\label{Zinfty}
\prob_{\psi_0}(Z=\infty) = 1-\int\limits_0^\infty dt \int\limits_{\bou} d^2\vx \, \vn(\vx)\cdot \vj^{\psi_t}(\vx) = \lim_{t\to\infty} \|\psi_t\|^2 \,.
\ee

Among other things, in this paper we deduce from the Lumer-Phillips theorem \cite{P59,LP61,EN00} that \eqref{Schr} and \eqref{bc} define a unique, autonomous time evolution for $\psi$, provided $\kappa(\vx)\geq 0$, see Theorem~\ref{thm:S} below. (If $\kappa(\vx)<0$ then the boundary condition \eqref{bc} is not absorbing but emitting, that is, there is a current coming out of the boundary; in this case, we would not expect a unique autonomous time evolution of $\psi$ to exist. 
For boundary points $\vx$ with $\kappa(\vx)=0$ the boundary condition is a Neumann boundary condition and thus reflecting.)

As we will explain, it follows further that if $\kappa(\vx) \geq 0$ everywhere, then the probability distribution given by \eqref{probnjR} and \eqref{Zinfty} can be defined for every $\psi_0\in L^2(\R,\CCC)$, and can be expressed in terms of a POVM (positive-operator-valued measure). Also, we treat not only dimension 3, but directly the obvious generalization to any dimension $d\in\NNN$.

In the presence of more than one particle in $\R$, the wave function must be collapsed appropriately when the first particle reaches $\bou$ and triggers a detector, and we have developed and discussed the appropriate equations in \cite{detect-several}. The $N$-particle Schr\"odinger equation in $\R^N$ gets supplemented by the appropriate boundary condition on $\partial(\R^N)$, which is
\be\label{bcn}
\vn_i(\vx_i)\cdot \nabla_i \psi(\vx_1,\ldots,\vx_N) = i\,\kappa(\vx_i) \psi(\vx_1,\ldots,\vx_N)
\quad \text{when } \vx_i\in\bou\,.
\ee
Suppose that at time $T^1$, the first detector gets triggered, in fact at location $\vX^1$ by particle number $I^1$. Now particle number $I^1$ gets absorbed and removed from consideration, and the wave function replaced by the conditional wave function
\be\label{psi'}
\psi'(x') = \mathcal {N}\, \psi_{T^1}(x',\vx_{I^1}=\vX^1)
\ee
with $x'\in \R^{N-1}$ any configuration of the remaining $N-1$ particles and $\mathcal{N}$ the appropriate normalizing factor. If $\psi$ is symmetric or anti-symmetric under permutations (as it would have to be for identical particles) then so will be $\psi'$. The process now repeats according to the corresponding equations for $N-1$ particles. 

\RT{Our Theorem~\ref{thm:N} will show that this process and the joint distribution of all detection events are well defined. To this end,} we need to explain what exactly \eqref{psi'} means and why $\psi'$ is a well-defined vector in $L^2(\R^{N-1})$; the difficulty comes from the fact that a general element of $L^2(\R^N)$, such as $\psi_{T^1}$, does not have well-defined values on a set of measure 0, such as the set where $\vx_{I^1}=\vX^1$. This point will be addressed by Theorem~\ref{thm:collapse} and its proof.

As steps towards Theorem~\ref{thm:collapse}, we prove in Theorem~\ref{thm:cond} in Section~\ref{sec:cond} that conditional wave functions in general have a well-defined distribution, and provide in Theorem~\ref{thm:prod} the POVM for an experiment done on a conditional wave function. These theorems can be applied also in other contexts independently of absorbing boundary conditions.

The analogous question of existence of solutions arises for the Dirac equation instead of the Laplacian, together with a suitable absorbing boundary condition \cite{detect-dirac}; some results on this question are given in \cite{TZZ}, and we plan to investigate it further in a future work. We also leave open here, for the Laplacian, the case of \emph{unbounded} regions $\Omega$, and limit our theorems to bounded ones. For further discussion of the absorbing boundary rule, see \cite{detect-rule,detect-several,DBD20,detect-derive,GTZ24}. For an overview of other proposals for the detection time distribution in quantum mechanics, see \cite{ML00}. Boundary conditions are also used for defining zero-range interactions; concerning the existence of solutions, see, e.g., \cite{AGHKH88,FLTZ23}.

The remainder of this paper is structured as follows:
In Section~\ref{sec:results}, we describe our theorems about absorbing boundaries. In Sections~\ref{sec:proofS}--\ref{sec:several}, we give the proofs. Our theorems about conditional wave functions are stated and proven in Section~\ref{sec:cond}.

\section{Results}
\label{sec:results}

\subsection{Single Particle}

We simply write $L^2(\Omega)$ for $L^2(\Omega,\CCC)$.
Our first theorem expresses that the Schr\"odinger equation with boundary condition \eqref{bc} actually defines a unique time evolution; the theorem can be formulated as follows. 
\begin{thm}\label{thm:S}
Suppose that $d\in \mathbb{N}$, that $\Omega \subset \mathbb{R}^d$ is a bounded {\LFn  domain} with Lipschitz boundary $\partial \Omega$, that $\kappa:\bou\to[0,\infty)$ is in $L^\infty(\partial \Omega)$, and that $V:\Omega\to\RRR$ is in $L^\infty(\Omega)$. Then there exists a {\LFn unique} dense subspace {\LFn $D(H)\subset H^{3/2}(\Omega)$} such that
\begin{enumerate}
\item for $\psi\in D(H)$, $\nabla^2\psi$ is defined as an element of $L^2(\Omega)$, and $\psi\big|_{\partial\Omega}$ and $\partial_n\psi\big|_{\partial\Omega}$ are defined {\LFn (see Lemma \ref{Lipschitz restriction})} as elements of $L^2(\partial\Omega)$;
\item for $\psi\in D(H)$, $\partial_n\psi = i\kappa \psi$ on $\partial\Omega$;
\item for every $\psi_0\in D(H)$, the 
initial-boundary value problem 
\begin{equation}\label{Robin IBVP}
\left\{\begin{array}{rclr}
       i \hbar\partial_t \psi&=& \bigl(-\frac{\hbar^2}{2m}\nabla^2 + V\bigr) \psi \quad &\text{in } \Omega
       \\
       \psi &=&\psi_0 \quad &\text{at } t=0
       \\
       \partial_n \psi&=&i \kappa \psi \quad &\text{on } \partial \Omega
    \end{array}
    \right.
    \end{equation}
admits a unique global-in-time solution
\be\label{strongersolutionsense}
\psi\in C\bigl([0,\infty),D(H)\bigr) \cap C^1\bigl([0,\infty),L^2(\Omega)\bigr)\,.
\ee
\end{enumerate}
Moreover, the operator $H=-\tfrac{\hbar^2}{2m}\nabla^2 + V:D(H)\to L^2(\Omega)$ generates a strongly continuous contraction semigroup $W_t=\exp(-iHt/\hbar): L^2(\Omega)\to L^2(\Omega)$. Thus, for every $\psi_0 \in L^2(\Omega)$, there is a unique global-in-time solution
$\psi\in C\bigl( [0,\infty),L^2(\Omega) \bigr)$ in the sense of contraction semigroups, given by $\psi_t=W_t\psi_0$ (and the two senses of solution agree for $\psi_0\in D(H)$).
If, moreover, $\Omega$ has a $C^2$ boundary and $\kappa \in C^1(\partial\Omega, [0,\infty))$, then we can explicitly characterize
\begin{equation}\label{H def}
D(H)=\bigl\{\psi \in H^2(\Omega):   \partial_n \psi=i \kappa \psi \text{ on } \partial \Omega\bigr\} \,.
\end{equation}
\end{thm}

\bigskip

We give the proof in Section~\ref{sec:proofS}. Here, ``Lipschitz boundary'' means that each point in $\partial\Omega$ has a neighborhood $U$ such that $U\cap\partial\Omega$ is, in some Cartesian coordinate system, the graph of a Lipschitz function $f$, i.e., $U\cap\partial\Omega=\{\vx:x_d=f(x_1,\ldots,x_{d-1})\}$; a ``$C^2$ boundary'' is one for which $f$ is twice continuously differentiable; $H^2(\R)$ denotes the second Sobolev space of $\R$, i.e., the space of $\psi_0 \in L^2(\R)$ whose second distributional derivatives lie in $L^2(\R)$. 
The terminology ``contraction'' means that $\|W_t \psi\|\leq \|\psi\|$; ``semigroup'' means that $W_t W_s=W_{t+s}$ for $t,s\geq 0$ and $W_0=I$ (the identity operator); ``strongly continuous'' means that $\lim_{t\to 0} \|W_t \psi_0 - \psi_0\| = 0$ for every $\psi_0\in L^2(\R)$. Since $W_t$ is in general not unitary, $\|\psi_t\|$ is in general smaller than $\|\psi_0\|$ for $t>0$ and for $\|\psi_0\|=1$ has the physical meaning of
\be
\|\psi_t\|^2 = \prob_{\psi_0} (T>t)\,.
\ee
The spectrum of a contraction $W$ lies in the closed unit disk $\{z\in\CCC: |z|\leq 1\}$ in the complex plane; however, $W$ is not necessarily diagonalizable \RT{(where ``diagonal'' means ``a multiplication operator'')}. The generator $H$ of a contraction semigroup has spectrum in the lower half plane $\{z\in\CCC: \Im\, z\leq 0\}$; again, $H$ need not be diagonalizable. In the present case neither $W_t$ nor $H$ are unitarily diagonalizable (they are not \emph{normal} operators, i.e., do not commute with their adjoints), as we show in Remark~\ref{rem:diag} in Section~\ref{sec:rem}. At least in some cases, $H$ can be diagonalized, but the \RT{generalized} eigenfunctions are not mutually orthogonal \cite{detect-imaginary}.

Intuitively speaking, the key difference between the two regularity conditions on $\partial\Omega$, Lipschitz and $C^2$, is that Lipschitz {\LFn regularity of $\partial \Omega$} allows for edges and corners (such as for a cube) while $C^2$ does not. A need for the weaker Lipschitz regularity assumption arises from the $N$-particle case, in which the domain in configuration space is $\Omega^N$, which will have edges and corners even if $\Omega$ has none (like a sphere); that is, $\Omega^N$ will not have a $C^2$ boundary even if $\Omega$ does, but still a Lipschitz boundary.

\bigskip

The next question that arises is whether the probability distribution \eqref{probnjR}  is well defined for a general $\psi$. The difficulty comes from the fact that \eqref{probnjR} involves evaluating $\psi_t$ on the boundary $\bou$, and $\psi_t$ may fail to be continuous; since a general element $\psi_t$ in $L^2(\R)$ is an equivalence class of functions modulo arbitrary changes on a set of volume 0, and since $\bou$ has volume 0, it is not well defined what $\psi_t$ is on $\bou$. A solution to this problem can be summarized as follows.

\begin{cor}\label{cor:S}
Under the assumptions of Theorem~\ref{thm:S}, there is a POVM $E(\cdot)$ on $[0,\infty)\times \bou \cup\{\infty\}$ acting on $L^2(\R)$ such that the probability distribution
\be\label{probE}
\prob_{\psi_0}(Z\in \,\cdot\,) = \scp{\psi_0}{E(\cdot)|\psi_0}
\ee 
(defined for every $\psi_0\in L^2(\R)$ with $\|\psi_0\|=1$) agrees with \eqref{probnjR} and \eqref{Zinfty} with $d-1$ dimensional surface integrals for $\psi_0 \in D(H)$ (for which \eqref{probnjR} and \eqref{Zinfty} are well defined). $E(\cdot)$ has the property that for every $\psi_0\in L^2(\R)$, the restriction of the measure $\scp{\psi_0}{E(\cdot)|\psi_0}$ to $[0,\infty)\times \bou$ is absolutely continuous (i.e., possesses a density) relative to the measure $dt \, d^{d-1}\vx$.
\end{cor}

We have included a proof in Section~\ref{sec:proofcorS}, making use of a strategy of Werner \cite{Wer87}. 

\begin{rem}
    {\LFn Since $D(H) \subset
H^{3/2}(\Omega)$ for Lipschitz boundary and $\kappa\in L^\infty(\partial \Omega,[0,\infty))$ while
$D(H)\subset H^2(\Omega)$ for $C^2$ boundary and $\kappa \in
C^1(\partial \Omega,[0,\infty))$, we have that the regularity} of $\psi_t$ is lower in general for Lipschitz boundary and bounded $\kappa$ than for $C^2$ boundary and {\LFn $\kappa  \in C^1(\partial \Omega,[0,\infty))$}; even for initial data $\psi_0$ from the space $C_c^\infty(\Omega)$ of smooth functions with compact support (which vanish on $\partial\Omega$ because $\Omega$ is open and so the support cannot touch $\partial \Omega$), $\psi_t$ will generically reside only in $H^{3/2}(\Omega)$ for $t>0$. Of course, the $C^2$ case is also of interest because it allows for a more explicit characterization of $D(H)$.\hfill$\diamond$
\end{rem}

\subsection{Many Particles}

We now turn to the case of $N$ particles. Theorem~\ref{thm:S} yields, after we replace $\R\to \R^N$ and $d\to Nd$, a well-defined time evolution\footnote{We can also allow the particles to have different masses $m_i$, as Theorem~\ref{thm:S} remains true when different masses are introduced for different components of the vector $\vx$. But for simplicity, we take the masses to be equal.} up to the first detection event for any $\psi_0\in L^2(\R^N)$ and a POVM on $[0,\infty)\times \partial(\R^N)\cup \{\infty\}$ that we will denote by $\tilde{E}$. However, it is not $\tilde{E}$ that gets measured by the detectors but really only a certain marginal of it that we will denote by $E'$: that is because the point of arrival on $\partial(\R^N)$ is an $N$-particle configuration that includes not just the position $\vx_i\in\bou$ of the particle, say $i$, that first hits $\bou$, but also the positions $\vx_j\in\R$ of all other particles at that time, while only $\vx_i$ gets measured but not $\vx_j$.

In more detail, the boundary $\partial(\R^N)$ consists of $N$ faces $\overline{F_i}$ corresponding to $\vx_i$ lying on the boundary $\bou$ while the other $\vx_j$'s may remain in the interior of $\R$ (which is $\R$, as we take $\R$ to be an open set); more precisely,
\be
\partial(\R^N) = \bigcup_{i=1}^N F_i \cup \bigcup_{\substack{i,j=1 \\ i\neq j}}^N F_{ij}
\ee
with
\begin{align}
F_i&:=\bigl\{ (\vx_1,\ldots,\vx_N)\in \overline{\R}^N: \vx_i\in \bou, \vx_j\in\R~\forall j\neq i \bigr\} \label{Fdef} \\
F_{ij}&:=\bigl\{(\vx_1,\ldots,\vx_N)\in \overline{\R}^N: \vx_i\in \bou, \vx_j\in\bou \bigr\} \,.
\end{align}
Note that $F_{ij}$ (an edge between two faces) has measure 0 in $\partial(\R^N)$, that the $F_i$ are mutually disjoint, and each $F_i$ is disjoint from each $F_{jk}$. Given that $Z\neq \infty$, the outcome $(T,X)$ is by Corollary~\ref{cor:S} continuously distributed in $[0,\infty)\times \partial(\R^N)$, so $X$ lies with probability 1 in one of the $F_i$, say $F_I$. We write $(T^1,\vX^1,I^1)$ for $(T,\vX_I,I)$ and introduce, for easier notation, the permutation function $p:\cup_i F_i \to \bou\times\{1,\ldots,N\}\times \R^{N-1}$ given by
\be\label{pdef}
p(x)=\bigl(\vx_i,i,\vx_1,\ldots,\vx_{i-1},\vx_{i+1},\ldots,\vx_N\bigr)~~\text{for }x=(\vx_1,\ldots,\vx_N)\in F_i\,.
\ee 
Note that $p$ is bijective and measure-preserving.
We define the POVM $E'$ on $[0,\infty)\times \bou\times \{1,\ldots,N\}\cup\{\infty\}$ by
\begin{align}
E'(\{\infty\})&=\tilde{E}(\{\infty\})\label{E'def1}\\
E'(dt \times B)&=\tilde{E}\bigl(dt \times p^{-1}(B\times \R^{N-1})\bigr)\label{E'def2}
\end{align}
for any measurable $B\subseteq \bou\times\{1,\ldots,N\}$. That is, away from $\infty$, $E'$ is the marginal of $\tilde{E}$ that corresponds to ignoring the $\vx_j$ other than $\vx_i\in\bou$. 

For example, for $N=2$ particles in $d=1$ dimension with $\Omega=(-1,1)$, e.g., $p(-1,x)=(-1,1,x)$ and $p(x,-1)=(-1,2,x)$ for $-1<x<1$, and $E'(dt \times \{-1\} \times \{2\})=\tilde{E}(dt \times \Omega \times \{-1\})$ is the operator for computing the probability that particle 2 gets detected first, in fact at time $T^1 \in dt$ in the location $\vX^1=-1$. Notice that the permutation $(I^1,\ldots,I^N)$ of the particles according to the order of detection is random, but the mapping $p$ is not.

\begin{cor}\label{cor:nS}
Let $N\in\NNN$, and let $\R\subset\RRR^d$, $\kappa$, and $V:\R^N\to\RRR$ be as in Theorem~\ref{thm:S}. Then the $N$-particle Schr\"odinger equation \eqref{Schr} in $\R^N$ and boundary condition \eqref{bcn} define a contraction semigroup $(W_t)_{t\geq 0}$ on $L^2(\R^N)$ and a POVM $\tilde{E}$ on $[0,\infty)\times\partial(\R^N)\cup\{\infty\}$ as before. Furthermore, for any $\psi_0\in L^2(\R^N)$ with $\|\psi_0\|=1$ and conditionally on $Z\neq\infty$, the joint distribution of $T^1,\vX^1,I^1$ exists, is absolutely continuous, and is given by $\scp{\psi_0}{E'(\cdot)|\psi_0}/\scp{\psi_0}{I-E'(\{\infty\})|\psi_0}$ with $E'$ defined in \eqref{E'def1}, \eqref{E'def2}.
\end{cor}

Next, we construct the entire process of $N$ detections; the crucial step is to guarantee the existence of the collapsed wave function. We also introduce another piece of notation: when we remove a particle, it is relevant to keep track of which particles remain, and to this end we introduce an index set $\sI$ with $N$ elements for labeling the particles. The notation $\R^\sI$ means the set of all mappings from $\sI$ to $\R$, i.e., all configurations in $\R$ with labels from $\sI$; put differently, $\R^\sI$ is the same as $\R^N$ but with the components labeled by elemens of $\sI$ instead of $1,\ldots,N$. The initial wave function will be a function on $\R^\sI$, and it is clear what is meant by the Laplacian on $\R^\sI$. 

\RT{
\begin{thm}\label{thm:collapse}
Let $\psi_t\in L^2(\R^\sI)$ follow the $N$-particle evolution with boundary condition \eqref{bcn} under the assumptions of Theorem~\ref{thm:S}, and $\|\psi_0\|=1$. Given that $T^1<\infty$ and $I^1=i$, it has probability 1 that $\psi'$ as in \eqref{psi'} is a well defined element of $L^2(\R^{\sI'})$ with $\sI'=\sI\setminus\{i\}$.
\end{thm}

We now define the joint distribution of several detection events by iterating the procedure for $\psi'$. To this end, suppose that for every index set $\sJ\subseteq \sI$ we are given a bounded potential $V_{\sJ}:\R^{\sJ}\to\RRR$ (intended to apply whenever only the particles with labels in $\sJ$ are still around). A natural choice of $V_\sJ$ for $x\in\R^\sJ$ would be, for example,
\be
V_\sJ(x) = \sum_{i\in\sJ}e_i \, V_1(\vx_i) + \sum_{\substack{i,j\in\sJ\\i\neq j}}e_ie_j\, V_2(\vx_i-\vx_j)
\ee
with arbitrary constants $e_i$ such as charges. For $k=1,\ldots,N$, we describe the $k$-th detection event through $Z^k=(\Delta T^k,\vX^k,I^k)$ (or $Z^k=\infty$ if fewer than $k$ detections occur) with waiting times $\Delta T^k=T^{k}-T^{k-1}$ (where $T^0=0$), detected places $\vX^k$, and labels $I^k$, defined inductively in the following way starting from $T^1, \vX^1,I^1$ as above, $\sI^{0}:=\sI$, and $\psi^{0}_t:=\psi_t$: If $Z^k=\infty$, then $Z^{k+1}=\infty$; otherwise, given that $I^k=i$, let $\sI^{k}=\sI^{k-1}\setminus \{i\}$, let the collapsed wave function $\psi^k_t \in L^2(\Omega^{\sI^k})$ have initial condition at $T^k$
\be
\psi^{k}_{T^k}(x) = \mathcal{N} \, \psi^{k-1}_{T^k}(x, \vx_i=\vX^k)
\ee
(well defined by Theorem~\ref{thm:collapse}) and evolve for $t\geq T^k$ according to the contraction semigroup $W^k_{t-T^k}$ defined by the Schr\"odinger equation \eqref{Schr} in $\Omega^{\sI^k}$ with potential $V=V_{\sI^k}$ and boundary condition \eqref{bcn}. Let $E^k$ be the corresponding instantiation of $E'$ as in Corollary~\ref{cor:nS} on $L^2(\R^{\sI^k})$; the distribution of $Z^{k+1}$ is then given by $\langle \psi^k_{T^k}|E^k(\cdot)|\psi^k_{T^k}\rangle$.

\begin{thm}\label{thm:N}
Under the assumptions of Theorem~\ref{thm:collapse} for $\psi_0\in L^2(\R^\sI)$ with $\|\psi_0\|=1$, the joint distribution of $(Z^1,\ldots,Z^N)$ exists as a measure on $\bigl([0,\infty)\times \bou\times \sI \cup\{\infty\}\bigr)^N$, 
is locally absolutely continuous, and is defined by a POVM $E_\sI$ acting on $L^2(\R^\sI)$.
\end{thm}

Note that not every element of $\bigl([0,\infty)\times \bou\times \sI \cup\{\infty\}\bigr)^N$ can actually occur; for example, $Z^k=\infty$ entails $Z^{k+1}=\infty$, and every $i\in\sI$ can come up at most once in the sequence. 
}

\section{Proof of Theorem~\ref{thm:S}}
\label{sec:proofS}

Theorem~\ref{thm:S} is an application of the Lumer-Phillips theorem \cite{P59,LP61,EN00} for contraction semigroups, which is a variant of the Hille-Yosida theorem \cite{EN00}. For our purpose the following version of this theorem due to Phillips \cite{P59} is most convenient. 

\begin{thm}[Lumer-Phillips Theorem for Contraction Semigroups] \label{thm:LP}
    Let $H$ be a closed linear operator defined on a dense linear subspace $D(H)$ of a Hilbert space $\Hilbert$. Moreover, assume that $-iH$ is dissipative, i.e., that for all $\psi\in D(H)$,
    \be\label{HY}
        \Re\langle \psi,-iH \psi \rangle_{\Hilbert} \leq 0 \,,
    \ee
    and that $-iH$ admits no dissipative extensions {\LFn or equivalently $\text{Ran}(H-i)=\Hilbert$}. Then $-iH/\hbar$ generates a strongly continuous semigroup of contractions $W_t=\exp(-iHt/\hbar)$ that preserves the domain of $H$, i.e., $W_t:D(H)\to D(H)$ for all $t \geq 0$. A sufficient condition for a closed dissipative operator $-iH$ to admit no dissipative extensions is that $iH^*$ is dissipative as well. 
\end{thm}

Our goal is to prove that the linear operator defined by equation (\ref{H def}) satisfies the assumptions of the Lumer-Phillips Theorem on the Hilbert space $L^2(\R)$. We first treat $C^2$ boundaries and {\LFn $\kappa \in C^1(\partial \Omega,[0,\infty))$}, and afterward turn to the case of weaker assumptions on $\partial\Omega$ and $\kappa$.

\subsection{$C^2$ Boundaries {\LFn and $C^1$ Boundary Condition Coefficients}}

To begin with, we need to explain what it means to set $\partial_n \psi= i \kappa \psi$ on $\partial \Omega$. For generic $\psi \in L^2(\R)$, there is no unambiguous way to define the restriction of the wave function or its derivative to a set of measure zero. Luckily, there is a classical result that provides a unique restriction map for wave functions residing in $H^2(\Omega)$. 

\begin{lm}\label{Trace Definition}
    \cite[Theorem 8.3]{LM} {\LFn \cite[Theorem 1.5.1.2]{Gr}} For $\Omega\subset \mathbb{R}^d$ a bounded $C^2$ domain, the restriction map $\psi \mapsto \left(\psi\big{|}_{\partial \Omega}, \partial_n \psi \big{|}_{\partial \Omega} \right)$ defined for $\psi \in C^{\infty}(\Omega) \to H^{3/2}(\partial \Omega)\times H^{1/2}(\partial \Omega)$ admits a continuous extension, also denoted $\psi \mapsto \left(\psi\big{|}_{\partial \Omega}, \partial_n \psi \big{|}_{\partial \Omega} \right)$, that is surjective for $H^2(\Omega)\to H^{3/2}(\partial \Omega)\times H^{{1/2}}(\partial \Omega)$ and admits a continuous right inverse. 
\end{lm}

{\LFn Now, we introduce the minimal Hamiltonian operator as $H_0:=\bigl(-\frac{\hbar^2}{2m}\nabla^2 +V\bigr)\big{|}_{C_c^\infty(\Omega)}$.} The operator $-iH_0$ is skew-symmetric on its domain, and it is well known {\LFn \cite{P59}} that any dissipative extension of a skew-symmetric operator must be some restriction of the ``maximal" operator $-iH_0^*$, where $H_0^*$ is the operator adjoint of $H_0$. We are thus motivated to define the operator $H_\kappa$ 
\begin{equation}\label{Hk def}
D(H_\kappa):=\bigl\{\psi \in D(H_0^*):   \partial_n \psi=i \kappa \psi \text{ on } \partial \Omega\bigr\}, \quad {H}_\kappa \psi:=H_0^*\psi
\end{equation}
with the intention of later proving that {\LFn $D(H_\kappa)\subset H^{2}(\Omega)$} and that this extension of $H_0$ generates a $C_0$ contraction semigroup. However, this definition for $H_\kappa$ does not make sense unless we can define unambiguous restriction maps for wave functions residing in $D(H_0^*)$.

\begin{lm}\label{extended trace}
    \cite[Theorem 8.3.9 and Corollary 8.3.11]{JHS} {\LFn \cite[Chapter 2]{Gr}} For $\Omega\subset \mathbb{R}^d$ a bounded $C^2$ domain, the restriction map $\psi \mapsto \left(\psi\big{|}_{\partial \Omega}, \partial_n \psi \big{|}_{\partial \Omega} \right)$ defined for $\psi \in H^2(\Omega) \to H^{3/2}(\partial \Omega)\times H^{1/2}(\partial \Omega)$ admits a continuous extension  to $D(H_0^*)\to H^{-1/2}(\partial \Omega)\times H^{-3/2}(\partial \Omega)$. In addition, for $\psi \in D(H_0^*)$ and $v \in H^2(\Omega)$, we have the integration by parts formula
    \begin{equation}\label{int-by-parts}
       \langle v, -iH_0^* \psi \rangle\Hi+ \langle -iH_0^*v, \psi \rangle\Hi =\frac{i\hbar^2}{2m}\int_{\partial \Omega} d^{d-1}\vx ~\Bigl(v^*\,\partial_n \psi - (\partial_n v^*) \, \psi \Bigr)
    \end{equation}
    {\LFn where the integrations in equation (\ref{int-by-parts}) are defined formally using the dual space pairings $\langle\cdot,\cdot\rangle_{H^s(\partial \Omega)\times H^{-s}(\partial \Omega)}$.} Lastly, the kernels of these restriction operators are subsets of $H^2(\Omega)$, i.e., if  $\psi \in D(H_0^*)$ satisfies the Dirichlet boundary condition $\psi \big{|}_{\partial \Omega}=0$ or the Neumann boundary condition $\partial_n \psi \big{|}_{\partial \Omega}=0$ then $\psi \in H^2(\Omega)$. 
\end{lm}
For $D(H_\kappa)$ to be well-defined, we now only need that multiplication by $\kappa$ is well-defined from $H^{-1/2}(\partial \Omega)\to H^{-3/2}(\partial \Omega)$. From the product rule we know that multiplication by a $C^1$ function $\kappa$ on a compact set such as $\partial\Omega$ defines a bounded linear map from $H^1(\partial \Omega) \to H^1(\partial \Omega)$. Let $\kappa':H^{-1}(\partial \Omega)\to H^{-1}(\partial \Omega)$ denote the Banach space adjoint of multiplication by $\kappa$. We show that $\kappa'\xi$ is simply $\kappa \xi$ for $\xi \in L^2(\partial \Omega)$, since $\kappa'\xi$ acts on elements $\chi \in H^1(\partial \Omega)$ according to
\begin{equation}
   \langle \chi, \kappa' \xi\rangle_{H^{1}(\partial \Omega)\times H^{-1}(\partial \Omega)}=\langle \kappa\chi, \xi\rangle_{H^{1}(\partial \Omega)\times H^{-1}(\partial \Omega)}= \int_{\partial \Omega} d^{d-1}\vx ~(\kappa \chi)^*  \xi=\int_{\partial \Omega} d^{d-1}\vx ~\chi^* (\kappa \xi).
\end{equation}
Since $L^2(\partial \Omega)$ is dense in $H^{-1}(\partial \Omega)$, multiplication by $\kappa$ extends uniquely to the bounded operator $\kappa'$ on $H^{-1}(\partial \Omega)$. Interpolation (see, e.g., {\LFn \cite[Theorem 1.4.3.3]{Gr}} and \cite[Theorems 5.1 and 7.7]{LM}) then implies that multiplication by $\kappa$ defines bounded linear operators from $H^s(\partial \Omega)\to H^s(\partial \Omega)$ for $s \in [-1,1]$, in particular $\kappa:H^{-1/2}(\partial \Omega)\to H^{-1/2}(\partial \Omega)\hookrightarrow H^{-3/2}(\partial \Omega)$ is bounded.

Now that we have a rigorous meaning for $H_\kappa$, we can verify that it satisfies the conditions of the Lumer-Phillips theorem. To show $H_\kappa$ is closed it suffices to show $D(H_\kappa)$ is closed in the graph norm. Since $D(H_\kappa)$ is the kernel of a bounded linear operator $\psi \mapsto \left(\partial_n \psi \big{|}_{\partial \Omega}-i\kappa\psi\big{|}_{\partial \Omega}\right)$ which maps $D(H_0^*)\to H^{-3/2}(\partial \Omega)$, $D(H_\kappa)$ is closed with respect to the graph norm of $H_0^*$. But $H_\kappa$ is a restriction of $H_0^*$, so their graph norms are the same on this domain and it follows $H_\kappa$ is closed.
 
To show that {\LFn $D(H_\kappa) \subset H^2(\Omega)$}, let $\psi \in D(H_\kappa)$. Then $\partial_n \psi \big{|}_{\partial \Omega} = i\kappa \psi \big{|}_{\partial \Omega} \in H^{-1/2}(\partial \Omega)$. It is known \cite[Lemma 3.2]{GM} that $\psi \mapsto \partial_n \psi \big{|}_{\partial \Omega}$ is a surjective operator from $H^{1/2}(\Omega)\cap D(H_0^*)\to H^{-1}(\partial \Omega)$ (and also from $H^{3/2}(\Omega)\cap D(H_0^*)\to L^2(\partial \Omega)$), so there exists some $\phi \in H^{1/2}(\Omega)\cap D(H_0^*)$ such that $\partial_n (\psi-\phi) \big{|}_{\partial \Omega}=0$. So, {\LFn by Lemma \ref{extended trace},} $\psi-\phi$ $\in H^2(\Omega)$, which implies that $\psi \in H^{1/2}(\Omega)$. We may now repeat the same argument: $\partial_n \psi \big{|}_{\partial \Omega} = i\kappa \psi \big{|}_{\partial \Omega}  \in L^{2}({\LFn \partial} \Omega)$, so there exists some $\phi \in H^{3/2}(\Omega)\cap D(H_0^*)$ such that $\partial_n (\psi-\phi) \big{|}_{\partial \Omega}=0$, which returns $\psi \in H^{3/2}(\Omega)$. Repeating the argument once more but now applying Lemma \ref{Trace Definition} returns $\psi \in H^2(\Omega)$ as desired.

We may now show that {\LFn $H_\kappa$} is dissipative by applying the integration by parts formula \eqref{int-by-parts}. For $\psi \in D({\LFn H_\kappa})$, we have that
\begin{align}
       2\, \Re \langle \psi, -i{\LFn H_\kappa}\psi \rangle\Hi 
       &=2 \,\Re \langle \psi, -iH_0^*\psi \rangle\Hi \\
       &=\frac{i\hbar^2}{2m} \int_{\partial\Omega} d^{d-1}\vx ~\Bigl( \psi^*  \, \partial_n\psi-  (\partial_n \psi^*) \, \psi \Bigr)\\
       &=-\frac{\hbar^2}{m} \int_{\partial \Omega} d^{d-1}\vx ~ \kappa (\vx) \, |\psi|^2(\vx) \leq 0
\end{align}
since $\kappa(\vx) \geq 0$ on $\partial \Omega$. To prove $-i{\LFn H_\kappa}$ admits no dissipative extensions, it is sufficient to show that its adjoint $i{\LFn H_\kappa^*}$ is also dissipative. We prove this by deriving an explicit expression for $D({\LFn H_\kappa^*})$. We first note that since ${\LFn H_\kappa}$ extends the minimal operator $H_0$, its adjoint ${\LFn H_\kappa^*}$ is extended by the maximal operator $H_0^*$. By definition of the operator adjoint, $\phi \in D(H_0^*)$ is an element of $D({\LFn H_\kappa^*})$ if and only if
\begin{equation}
 \langle \psi, -iH_0^* \phi \rangle\Hi+\langle -i{\LFn H_\kappa}\psi, \phi \rangle\Hi =0
\end{equation}
for all $\psi \in D({\LFn H_\kappa})$. Applying the integration by parts formula, we see that this holds if and only if
\begin{equation}
\int_{\partial \Omega} d^{d-1}\vx ~\Bigl( \psi^* \, \partial_n \phi - (\partial_n \psi^*) \, \phi\Bigr)= \int_{\partial \Omega} d^{d-1}\vx \,\psi^* \, (\partial_n \phi +i\kappa \phi)=0
\end{equation}
for all $\psi \in D({\LFn H_\kappa})$. Split surjectivity of the restriction operators implies that for every $\xi \in H^{3/2}(\partial \Omega)$ there exists a wave function $\psi \in H^2(\Omega)$ with $(\psi \big{|}_{\partial \Omega}, \partial_n \psi \big{|}_{\partial \Omega})= (\xi, i \kappa \xi)$. In other words, for every $\xi \in H^{3/2}(\partial \Omega)$ there exists a $\psi \in D({\LFn H_\kappa})$ with $\psi \big{|}_{\partial \Omega}= \xi$. It follows that $\phi \in D({\LFn H_\kappa^*})$ if and only if 
\begin{equation}
\int_{\partial \Omega} d^{d-1}\vx \,\xi^* \, (\partial_n \phi +i\kappa \phi) =0
\end{equation}
for all $\xi \in H^{3/2}(\partial \Omega)$, which is a dense subspace, so the domain of the adjoint must be given by
\begin{equation}\label{DH*}
D({\LFn H_\kappa^*})=\bigl\{\phi \in D(H_0^*):   \partial_n \phi=-i \kappa \phi \text{ on } \partial \Omega\bigr\}.
\end{equation}
One can now repeat the same steps to prove $D({\LFn H_\kappa^*})\subset H^{2}(\Omega)$ and apply the integration by parts formula to show dissipativity of $i{\LFn H_\kappa^*}$. This completes the proof of Theorem~\ref{thm:S} for $C^2$ boundaries and {\LFn $\kappa \in C^1(\partial \Omega,[0,\infty))$}. We now turn to the case of weaker assumptions.

\subsection{Lipschitz Boundaries {\LFn and $L^\infty$ Boundary Condition Coefficients}}

The extension to the Lipschitz boundary case requires a significant amount of technical machinery. Fortunately, the theory of closed extensions of $(-\nabla^2 + V)\big{|}_{C^\infty_c(\Omega)}$ on Lipschitz domains has received a lot of attention in recent years, and the result that we seek {\LFn has effectively already been proven using the framework of \textit{quasi boundary triples}. The results from this framework are directly applicable to our setting, but their proofs are quite technical so we will primarily refer the reader to \cite{JL07,GM,JM} for further reading.
\begin{definition}
    Let $H_0$ be a densely defined closed symmetric operator in a Hilbert space $\Hilbert$ and assume that $T$ is a linear operator in $\Hilbert$ such that $\overline{T}=H^*_0$. A \textbf{quasi boundary triple} for $T$ consists of another Hilbert space $\mathcal{G}$ and two linear mappings $\Gamma_0,\Gamma_1:D(T)\to\mathcal{G}$ satisfying
    \begin{enumerate}
        \item For all $v,\psi \in D(T)$, the abstract Green's identity holds
    \begin{equation}\label{abstract Green}
        \langle v, T\psi \rangle_{\Hilbert}-\langle Tv,\psi \rangle_{\Hilbert}= \langle \Gamma_0 v, \Gamma_1 \psi\rangle_{\mathcal{G}}-\langle \Gamma_1 v, \Gamma_0 \psi \rangle_{\mathcal{G}}
    \end{equation}
    \item $(\Gamma_0,\Gamma_1):D(T)\to \mathcal{G}\times \mathcal{G}$ has dense range
    \item The operator $A_0:=T\big{|}_{\ker(\Gamma_0)}$ is self-adjoint in $\Hilbert$.
    \end{enumerate}
\end{definition}
For $\lambda \in \rho(A_0)$, the self-adjointness of $A_0$ provides a natural decomposition of the domain of $T$ into $D(T)=D(A_0)\oplus \ker(T-\lambda)=\ker(\Gamma_0)\oplus \ker(T-\lambda)$. Hence for every $\xi \in \text{Ran}(\Gamma_0)$ there exists a unique solution $\left(\Gamma_0\big{|}_{\ker(T-\lambda)} \right)^{-1}\xi:= \phi_\lambda$ of the abstract boundary value problem
\begin{equation}
    (T-\lambda)\phi_\lambda=0, \quad \Gamma_0 \phi_\lambda=\xi.
\end{equation}
We define for $\lambda \in \rho(A_0)$ the \textbf{Weyl function} $M(\lambda)$ associated to the triple $\{\mathcal{G},\Gamma_0,\Gamma_1\}$ as the linear mapping
    \begin{equation}
        M(\lambda):=\Gamma_1\left(\left(\Gamma_0\big{|}_{\ker(T-\lambda)} \right)^{-1}\right):\text{Ran}(\Gamma_0)\to\text{Ran}(\Gamma_1).
    \end{equation}
    Under certain circumstances (which occur frequently for physical applications), the quasi boundary triple framework provides a simple set of conditions that one can check to prove that a large class of restrictions of $-iT$ generate $C_0$ contraction semigroups. For linear mappings $\kappa:\mathcal{G}\to\text{Ran}(\Gamma_1)\subset \mathcal{G}$, we will be primarily interested in restrictions $A_\kappa$ of the form
\begin{equation}\label{A k def}
            D(A_\kappa):=\{ \psi \in D(T): \Gamma_1\psi=-i\kappa \Gamma_0\psi\}, \quad A_\kappa\psi:=T\psi.
        \end{equation}
\begin{lm}\label{Krein}
\cite[Theorem 2.8]{JL07}
        Let $H_0$ be a densely defined closed symmetric operator in $\Hilbert$ and let $\{\mathcal{G},\Gamma_0,\Gamma_1\}$ be a quasi boundary triple for $\overline{T}= H_0^*$. If $\kappa:\mathcal{G}\to \text{Ran}(\Gamma_1)\subset\mathcal{G}$ is a bounded operator on $\mathcal{G}$ such that $i\kappa+M(i)$ is injective and surjective onto $\text{Ran}(\Gamma_1)$, then $\text{Ran}(A_\kappa-i)=\Hilbert$.
    \end{lm}

Now, for our setting the symmetric} \textit{minimal operator} {\LFn  is} $H_0:= \left(-\tfrac{\hbar^2}{2m}\nabla^2 + V \right)\big{|}_{H^2_0(\Omega)}$. The relevant notion of restriction to Lipschitz boundaries $\partial \Omega$ {\LFn is given in the following lemma.}

\begin{lm}\label{Lipschitz restriction}
    \cite[Lemma 3.1 and Lemma 3.2]{GM} For $\Omega \subset \mathbb{R}^d$ a bounded Lipschitz domain, the restriction maps $\psi \mapsto \left(\psi\big{|}_{\partial \Omega}, \partial_n \psi \big{|}_{\partial \Omega} \right)$ defined for $\psi \in C^{\infty}(\Omega) \to H^{1}(\partial \Omega)\times L^2(\partial \Omega)$ admits continuous {\LFn surjective extensions $\psi \mapsto \psi\big{|}_{\partial \Omega}$ for $D(H_0^*)\cap H^{3/2}(\Omega)\to H^{1}(\partial \Omega)$ and $\psi \mapsto \partial_n\psi\big{|}_{\partial \Omega}$ for $D(H_0^*)\cap H^{3/2}(\Omega)\to L^{2}(\partial \Omega)$. }
\end{lm}
{\LFn These restriction maps allow us to construct a quasi boundary triple for an extension of $H_0$.}
{\LFn \begin{lm}\label{QBT construction}
 \cite[Theorem 4.1]{JM} Let $\Omega\subset \mathbb{R}^d$ be a bounded Lipschitz domain and define the linear operator $T$ via
 \begin{equation}
     D(T):=D(H_0^*)\cap H^{3/2}(\Omega), \quad T=H_0^*\big{|}_{D(T)}.
 \end{equation}
 Then $\{L^2(\partial \Omega), \Gamma_0,\Gamma_1 \}$ form a quasi boundary triple for $T\subset H_0^*$ where 
 \begin{equation}
     \Gamma_0\psi:=\frac{\hbar}{\sqrt{2m}}\psi \big{|}_{\partial \Omega}, \quad \Gamma_1\psi:=-\frac{\hbar}{\sqrt{2m}}\partial_n \psi \big{|}_{\partial \Omega}.
 \end{equation}
 In addition, the \textit{Dirichlet} and \textit{Neumann} Hamiltonians
 \begin{equation}
     H_D:=T \big{|}_{\ker(\Gamma_0)}, \quad H_N=T\big{|}_{\ker(\Gamma_1)}
 \end{equation}
 are both self-adjoint operators in $L^2(\Omega)$. Lastly, for $\lambda \in \rho(H_D)\cap\rho(H_N)$ the Weyl mapping $M(\lambda):H^1(\partial \Omega)\to L^{2}(\partial \Omega)$ is a bounded invertible operator which takes solutions of the Dirichlet boundary value problem
 \begin{equation}
     \left(-\frac{\hbar^2}{2m}\nabla^2 + V \right)\psi_\lambda=\lambda \psi_\lambda, \quad \frac{\hbar}{\sqrt{2m}}\psi_\lambda \big{|}_{\partial \Omega}=\xi\in L^2(\partial \Omega)
 \end{equation}
 and maps them to their negative Neumann boundary values $M(\lambda):\xi \mapsto \frac{-\hbar}{\sqrt{2m}}\partial_n\psi_\lambda \big{|}_{\partial \Omega}$.
\end{lm}
Theorem~\ref{thm:S} for Lipschitz boundaries $\partial \Omega$ and bounded coefficients $\kappa \in L^\infty(\partial \Omega,[0,\infty))$ follows immediately from the Corollary below.
\begin{cor}
    Let $\Omega\subset \mathbb{R}^d$ be a bounded Lipschitz domain and $\kappa \in L^\infty(\partial \Omega,[0,\infty))$. Then the operator $-iH_\kappa$ defined by
    \begin{equation}
D(H_\kappa):=\bigl\{\psi \in D(H_0^*)\cap H^{3/2}(\Omega):   \partial_n \psi=i \kappa \psi \text{ on } \partial \Omega\bigr\}, \quad H_\kappa\psi:=H_0^*\psi.
\end{equation}
is a dissipative extension of $-iH_0$ satisfying $\text{Ran}(H_\kappa-i)=\Hilbert$. Hence, $-i H_\kappa / \hbar$ generates a strongly continuous semigroup of contractions $W_t= \exp(-iH_\kappa t/\hbar)$ on $L^2(\Omega)$ that preserves the domain of $H_\kappa$\em, i.e., $W_t:D(H_\kappa) \to D(H_\kappa)$ for all $t \geq 0$. 
\end{cor}
\begin{proof}
    By Lemma \ref{Krein}, Lemma \ref{Lipschitz restriction}, and Lemma \ref{QBT construction} it is sufficient to prove that $i\kappa + M(i):H^1(\partial \Omega)\to L^2(\partial \Omega)$ is a bijective linear operator. To prove that $i\kappa + M(i)$ is injective we compute for $\xi \in H^1(\partial \Omega)$
    \begin{equation}\label{injectivity}
        \Im\langle \xi, \left(i \kappa + M(i)\right)\xi\rangle_{L^2(\partial \Omega)}\geq \Im\langle \xi,  M(i)\xi\rangle_{L^2(\partial \Omega)}= || \left(\Gamma_0 \big{|}_{\ker(T-i)} \right)^{-1}\xi||^2_{\mathcal{H}}>0
    \end{equation}
    where the equality in (\ref{injectivity}) follows from the abstract Green's formula (\ref{abstract Green}). Since $H_N$ is self-adjoint we have $i \in \rho(H_N)$ and $M(i)^{-1}:L^2(\partial \Omega)\to H^1(\partial \Omega)$ is a compact operator on $L^2(\partial \Omega)$ by Rellich-Kondrachov Theorem \cite[Theorem 6.3]{Adams}. The injectivity of $i\kappa + M(i)$ along with the bijectivity of $M(i)$ imply that the operator $i\kappa M(i)^{-1}+I$ is also an injective operator on $L^2(\partial \Omega)$. Now, $i \kappa M(i)^{-1}$ is a compact operator on $L^2(\partial \Omega)$, so the Fredholm alternative \cite[Theorem 6.6]{Brezis} states that the injective compact perturbation of the identity $i\kappa M(i)^{-1}+I$ is bijective. Hence $i\kappa + M(i)$ is also bijective, concluding our proof.
\end{proof}}

Now all statements of Theorem~\ref{thm:S} follow.

\section{Remarks}
\label{sec:rem}

\begin{rem}
    For $\psi_0 \in D(H)$, the Lumer-Phillips theorem provides the existence of a solution $\psi_t \in {\LFn D(H)}$ for $t \geq 0$ to the initial boundary value problem (\ref{Robin IBVP}). The uniqueness of this solution is straightforward. Let $\psi_0 \in {\LFn D(H)}$, $\psi_t' \in {\LFn D(H)}$ be a solution to (\ref{Robin IBVP}) for $t \geq 0$. By linearity $\phi_t:= W_t\psi_0 - \psi_t'$ solves (\ref{Robin IBVP}) with initial condition $\phi_0=0$. But the boundary condition implies $\frac{d}{dt}||\phi_t||_{L^2(\Omega)}\leq 0$ for all $t \geq 0$, hence $\phi_t=0$ for all time and the two solutions are equal.
    \par
    {\LFn A similar line of reasoning shows that the operator $H_\kappa$ satisfying the three conditions of Theorem \ref{thm:S} is unique. Suppose there was another operator $H$ with domain $D(H)\subset H^{3/2}(\Omega)$ satisfying the properties outlined in Theorem \ref{thm:S}. Let $\psi_0 \in D(H)$ and let $\psi_0^n \in D(H_k)$ be a sequence that approximates $\psi_0$ in $L^2(\Omega)$ norm. Then $\phi_t^n:=\exp(-itH)\psi_0 - \exp(-itH_\kappa)\psi_0^n$ satisfies (\ref{Robin IBVP}), and consequently $||\phi_t^n||_{L^2(\Omega)}\leq ||\phi_0^n||_{L^2(\Omega)}$. Hence $\lim_{n \to \infty}||\exp(-itH)\psi_0 - \exp(-itH_\kappa) \psi_0^n||_{L^2(\Omega)}=0$, in particular $\exp(-itH)\psi_0=\lim_{n \to \infty}\exp(-itH_\kappa)\psi^n_0=\exp(-itH_\kappa)\psi_0$. Differentiating this returns $\frac{d}{dt}\exp(-itH_k)\psi_0\big{|}_{t=0}=H\psi_0\in L^2(\Omega)$, so the Hille-Yosida theorem \cite{EN00} states that $\psi_0$ must reside in the domain of the generator $H_\kappa$. It follows that $D(H)\subset D(H_\kappa)$, and repeating the same steps with $\psi_0\in D(H_\kappa)$ returns $D(H_\kappa)\subset D(H)$.} \hfill$\diamond$
\end{rem}

\begin{rem}
Suppose we replace $i\kappa$ in the boundary condition \eqref{bc} by $\nu+i\kappa$ with $\nu \in {\LFn  L^\infty(\partial \Omega, \RRR)}$, so that \eqref{bc} becomes
\be\label{bc2}
\frac{\partial \psi}{\partial n}(\vx) = (\nu(\vx)+i\kappa(\vx)) \psi(\vx)\,.
\ee
Then this boundary condition is still absorbing, i.e., one that forces the current to point outward. One can repeat the proof of Theorem~\ref{thm:S} with this boundary condition  to show this also generates a contraction semigroup.\hfill$\diamond$
\end{rem}

\begin{rem}\label{rem:diag} 
Unlike self-adjoint Hamiltonians, $H$ is not unitarily diagonalizable when $\kappa(\vx)>0$ on a set of $\vx$s of positive measure in $\bou$, as we prove below. (We note also that the Hamiltonian of the discrete version of the absorbing boundary rule for a quantum particle on a lattice is easily checked to be non-normal ($HH^*\neq H^*H$), and thus not unitarily diagonalizable.) It seems that, at least in many cases, a complete set of (generalized, non-normalizable) eigenfunctions exists, but they are not mutually orthogonal \cite{detect-imaginary}.

Recall that an operator $A$ in $\Hilbert$ is unitarily diagonalizable if and only if there is a generalized orthonormal basis, i.e., a unitary isomorphism $U:\Hilbert\to L^2(S,\mu)$ for some measure space $(S,\mu)$, such that $M=UAU^{-1}$ is the multiplication operator by some function $f:S\to \CCC$. The domain $D(M)$ on which the graph of $M$ is closed is given by
\be
D(M)=\Bigl\{\psi\in L^2(S,\mu): \int_S |f(s) \, \psi(s)|^2 \, \mu(ds) < \infty\Bigr\}\,.
\ee
Since the adjoint $T^*$ of any operator $T$ with domain $D(T)$ is defined on the domain
\be
D(T^*) = \Bigl\{ \psi\in\Hilbert: \exists \phi\in\Hilbert:\forall \chi\in D(T):  \scp{\psi}{T\chi} = \scp{\phi}{\chi}  \Bigr\}
\ee
(and given there by $T^*\psi=\phi$), the adjoint $M^*$ of a multiplication operator $M$ has domain $D(M^*) = D(M)$ and is given there by multiplication by $f^*$. When $\kappa(\vx)>0$ on a set of positive measure, then, as the proof of Theorem~\ref{thm:S} has shown, $H$ has domain $D(H)$ different from $D(H^*)$, see \eqref{DH*}, while the graph of $H$ is closed, so it follows that $H$ cannot be unitarily diagonalizable.\hfill$\diamond$
\end{rem}

\section{Proof of Corollary~\ref{cor:S}}
\label{sec:proofcorS}

For any $\psi_0\in D(H)$, also $\psi_t=\exp(-iHt/\hbar)\psi_0$ lies in $D(H)$. Moreover, for any $\psi\in D(H)$, $\vn(\vx)\cdot \vj^\psi(\vx)= (\hbar\kappa(\vx)/m) |\psi(\vx)|^2$ on $\bou$, and the restriction of $\psi$ to $\bou$ is well defined as an element of $L^2(\bou,d^{d-1}\vx)$ . It follows that for $\psi_0\in D(H)$ with $\|\psi_0\|=1$, \eqref{probnjR} and \eqref{Zinfty} together define a probability distribution on $[0,\infty) \times \bou \cup\{\infty\}$.

Now define, for $\psi_0\in D(H)$, $J\psi_0$ to be the function on $[0,\infty)\times \bou$ such that $J\psi_0(t,\cdot)$ is $\sqrt{\hbar\kappa(\vx)/m}$ times the restriction of $\psi_t$ to $\bou$. Then $J\psi_0 \in L^2 \bigl( [0,\infty)\times \bou, dt\, d^{d-1}\vx \bigr)$, and
\be\label{Jbound}
\|J\psi_0\|^2 = \frac{\hbar}{m} \int_0^\infty dt \int_{\bou} d^{d-1}\vx \, \kappa(\vx)\,|\psi_t(\vx)|^2 = \|\psi_0\|^2 -\lim_{t\to\infty} \|\psi_t\|^2 \leq \|\psi_0\|^2\,.
\ee
(Note that $\lim_{t\to\infty} \|\psi_t\|^2$ exists because $t\mapsto \|\psi_t\|^2$ is a non-negative, decreasing function.) The fact $\|J\psi_0\|\leq \|\psi_0\|$ means that $J: D(H)\to L^2\bigl( [0,\infty)\times \bou, dt\, d^{d-1}\vx \bigr)$ is a bounded operator with operator norm no greater than 1 (i.e., a contraction). Thus, $J$ possesses a unique bounded extension to $L^2(\R)$, which we will also denote by $J$. 

For arbitrary $\psi_0\in L^2(\R)$ (outside $D(H)$) with $\|\psi_0\|=1$, $|J\psi_0(t,\vx)|^2$ is the joint probability density of $T$ and $\vX$, and $1-\|J\psi_0\|^2=\prob_{\psi_0}(Z=\infty)$; that is, the distribution of $Z$ is well defined. The POVM $E$ is given on $[0,\infty)\times \bou$ by
\be
E(\cdot)= J^*P(\cdot)J\,,
\ee
where $P$ is the natural PVM (projection-valued measure) on $L^2 \bigl( [0,\infty)\times \bou, dt\, d^{d-1}\vx \bigr)$. (The \emph{natural PVM} on $L^2(S,\mu)$ associates with every measurable subset $B$ of a measure space $(S,\mu)$ the projection to the subspace consisting of the functions vanishing outside $B$.) The definition of $E$ is completed by setting $E(\{\infty\})=I-J^*J$, which is a positive operator by \eqref{Jbound}. It follows that $E \bigl( [0,\infty)\times \bou\cup\{\infty\} \bigr)=I$, so $E$ is a POVM, and that \eqref{probE} agrees with \eqref{probnjR} and \eqref{Zinfty} for $\psi_0\in D(H)$. It also follows that $E(\{\infty\}) = \lim_{t\to\infty} W_t^* W_t$ because $W_t^*W_t = E([t,\infty)\times \bou\cup\{\infty\})$. 

Concerning the last sentence of Corollary~\ref{cor:S}, the absolute continuity of the measure $\scp{\psi_0}{E(\cdot)|\psi_0}$ on $[0,\infty)\times \bou$ is visible from the fact that $|J\psi_0(t,\vx)|^2$ is its density.

\section{Several Particles}
\label{sec:several}

The main new issue about the case of several particles is whether the collapsed wave function $\psi'$ in \eqref{psi'} is well defined. To this end, we begin with some general considerations about conditional wave functions.

\subsection{Conditional Wave Functions}
\label{sec:cond}

Let $\SSS(\Hilbert)$ denote the unit sphere of the Hilbert space $\Hilbert$, which we consider with its Borel $\sigma$-algebra.

In general, for a wave function $\psi(a,b)$ of two variables $a,b$, the conditional wave function $\psi'$ is defined as follows: insert for $a$ a random value $A$ whose distribution is the appropriate marginal of $|\psi|^2$, and then normalize. Thus, $\psi'$ is a random function of the single variable $b$. More generally, we can consider a function $\psi(a)$ with values in some Hilbert space $\Hilbert_b$ (including the special case that $\Hilbert_b$ are the $L^2$ functions of the variable $b$), and then we want that $\psi'$ is a random variable with values in $\SSS(\Hilbert_b)$. One question that arises is whether $\psi(A)$ might be 0; we show that this happens with probability 0. Another question arises from the fact that an element of an $L^2$ space is strictly speaking not a function but an equivalence class of functions that can differ on a set of measure 0; we show that $\psi'$ is ``almost uniquely'' defined. This is done in the next theorem. 

\begin{thm}\label{thm:cond}
Let $\sA$ be a measure space such that $L^2(\sA)$ is separable, $\Hilbert_b$ another separable Hilbert space, $\psi\in L^2(\sA,\Hilbert_b)=L^2(\sA)\otimes \Hilbert_b$ with $\|\psi\|=1$, $\tilde\psi:\sA\to\Hilbert_b$ a representative  of $\psi$, and $A$ an $\sA$-valued random variable with distribution density $\|\tilde\psi(a)\|_b^2$. Then $\psi':=\tilde\psi(A)/\|\tilde\psi(A)\|_b$ is almost surely well defined as an element of $\SSS(\Hilbert_b)$, and the probability distribution of the pair $(A,\psi')$ in $\sA\times \SSS(\Hilbert_b)$ does not depend on the choice of $\tilde\psi$.
\end{thm}

\begin{proof}
We first verify that $L^2(\sA,\Hilbert_b)=L^2(\sA)\otimes \Hilbert_b$ in the sense that they are canonically isomorphic; the argument is a variant of one in \cite[p.~51]{RS1}. Given ONBs $\{\alpha_i\}$ of $L^2(\sA)$ and $\{\beta_j\}$ of $\Hilbert_b$, the functions $\gamma_{ij}(a)=\alpha_i(a) \, \beta_j$ lie in $L^2(\sA,\Hilbert_b)$ and are orthonormal. To see that they form an ONB, suppose that $f\in L^2(\sA,\Hilbert_b)$ and
\begin{align}
0&=\scp{f}{\gamma_{ij}}_{L^2(\sA,\Hilbert_b)}\\
&= \int_{\sA} da \, \scp{f(a)}{\gamma_{ij}(a)}_b\\
&= \int_{\sA} da \, \alpha_i(a) \scp{f(a)}{\beta_j}_b\,.
\end{align}
Since $\{\alpha_i\}$ is an ONB, it follows that $0=\scp{f(a)}{\beta_j}_b$ for almost every $a$, and since $\{\beta_j\}$ is an ONB and countable that $f(a)=0$ for almost every $a$.

Therefore, $\gamma_{ij} \mapsto \alpha_i \otimes \beta_j$ maps an ONB to an ONB, so its unique continuous linear extension is a unitary isomorphism from $L^2(\sA,\Hilbert_b)$ to $L^2(\sA)\otimes \Hilbert_b$.

Now pick any function $\tilde\psi$ belonging to the equivalence class of functions that $\psi$ is and let $A$ be a random variable taking values in $\sA$ with $|\tilde\psi|^2$ distribution, i.e.,
\be\label{Adistribution}
\prob(A\in S) = \int_{S}da\, \|\tilde\psi(a)\|^2_b
\ee
for all measurable subsets $S$ of $\sA$. 
For the purpose of inserting $A$, we first leave normalization aside 
and set $\psi_*:=\tilde\psi(A)$. If we had picked another function $\hat\psi$ instead of $\tilde\psi$, then $\hat\psi$ would differ from $\tilde\psi$ on a set of measure 0 in $\sA$. Thus, the distribution \eqref{Adistribution} of $A$ is independent of whether we choose $\tilde\psi$ or $\hat\psi$, and $\hat\psi(A)=\tilde\psi(A)$ with probability 1. Thus, the distribution of $(A,\psi_*)$ is a well-defined probability measure in $\sA\times \Hilbert_b$.

Next we focus on normalization: $A$ has probability 1 to be such that the norm in $\Hilbert_b$ of $\psi_*$ is non-zero. After all, $A$ has probability 0 by \eqref{Adistribution} to lie in the set of $a$ values with $\|\tilde\psi(a)\|^2_b=0$. Thus, $\psi_*$ can be normalized, i.e., $\mathcal{N}:=1/\|\psi_*\|$ and $\psi' = \mathcal{N}\psi_*$ exists. Since in any Hilbert space $\Hilbert$ the normalization mapping $\Hilbert\setminus\{0\} \to \SSS(\Hilbert)$, $\phi\mapsto \phi/\|\phi\|$ is continuous, it is Borel-measurable. Thus, the distribution of $(A,\psi')$ is defined on $\sA\times \SSS(\Hilbert_b)$ and is independent of the choice of $\tilde\psi$ within the equivalence class that is $\psi$. 
\end{proof}

\begin{rem}\label{rem:2}
For $\Hilbert_b=L^2(\sB)$ for some measure space $\sB$, we obtain, since $L^2(\sA)\otimes L^2(\sB)=L^2(\sA\times \sB)$ with the product measure, that from an $L^2$ function $\psi(a,b)$ we can form the conditional wave function $\psi'(b)=\mathcal{N} \, \psi(A,b)$, and the pair $(A,\psi')$ has a well-defined probability distribution in $\sA\times \SSS(L^2(\sB))$.\hfill$\diamond$
\end{rem}

\begin{rem}\label{rem:3}
For a function $\psi(a,b,c)$ of three variables, first conditioning on the $a$ variable (i.e., inserting $A$ and normalizing) to obtain $\psi'(b,c)$ and then conditioning on the $b$ variable to obtain $\psi''(c)$ yields the same distribution for $(A,B,\psi'')$ as first conditioning on the $b$ variable and then on $a$, and the same distribution as conditioning on the pair $(a,b)$.\hfill$\diamond$
\end{rem}

\begin{rem}\label{rem:bundle}
Theorem~\ref{thm:cond} remains true when we replace the fixed Hilbert space $\Hilbert_b$ by a measurable bundle $\Hilbert_b(a)$ of Hilbert spaces, that is, if $\Hilbert_b$ depends on $a\in\sA$; then (a representative of) $\psi$ is a measurable cross-section of this bundle.\hfill$\diamond$
\end{rem}

The next theorem is concerned with an experiment with POVM $E$ done on a conditional wave function and asserts that the joint distribution of the outcome $B$ of the experiment and the $A$ on which we conditioned is defined by a product POVM. We begin with the definition of the latter: For any two POVMs $E,F$ on measurable spaces $\sA,\sB$ acting on Hilbert spaces $\Hilbert_a,\Hilbert_b$, a \emph{product POVM} is a POVM $G$ on $\sA\times \sB$ acting on $\Hilbert_a\otimes \Hilbert_b$ such that for any measurable subsets $A,B$,
\be\label{productPOVM}
G(A\times B) = E(A) \otimes F(B)\,.
\ee
Note that on subsets $C$ of $\sA\times\sB$, $G(C)$ does not have to be a tensor product. If $G$ exists and is unique, we write $E\otimes F$ for $G$.

\begin{prop}\label{prop:uniqueprod}
The product POVM is unique if $\Hilbert_a$ and $\Hilbert_b$ are separable.
\end{prop}

\begin{proof}
For $\alpha\in \Hilbert_a$ and $\beta\in\Hilbert_b$,
\be
\scp{\alpha\otimes\beta}{E(A) \otimes F(B)|\alpha\otimes \beta}= \scp{\alpha}{E(A)|\alpha}_a ~ \scp{\beta}{F(B)|\beta}_b
\ee
is a product of finite measures and thus always extends uniquely to a measure on $\sA\times \sB$. Likewise for $\alpha'\neq \alpha$ and $\beta'\neq \beta$, $\scp{\alpha'\otimes\beta'}{E(A) \otimes F(B)|\alpha\otimes \beta}$ extends uniquely to a complex measure on $\sA\times \sB$. 
By separability, $\psi\in\Hilbert_a\otimes\Hilbert_b$ can be written as a countable series $\psi=\sum_{ij}c_{ij}\alpha_i\otimes\beta_j$ using ONBs, so the complex measure $\scp{\psi'}{G(C)|\psi}$ can be expanded into a convergent series of complex measures and therefore is determined uniquely (while the convergence of the series would not be obvious if nothing is known about the existence of $G$).
\end{proof}

We conjecture that $G$ exists for any two POVMs, but as far as we know this has been proved only under the additional assumption that $\sA$ and $\sB$ are standard Borel spaces,\footnote{A measurable space is called a \emph{standard Borel space} if and only if it is isomorphic as a measurable space to a complete separable metric space with its Borel $\sigma$-algebra. This is not a strong restriction as it includes most spaces that one considers in practice, such as countable unions of closed subsets of manifolds.} see Corollary 7 in \cite{crea2A}. However, we do not need this assumption for our result:

\begin{thm}\label{thm:prod}
Let $\sA$ be a measure space such that $L^2(\sA)$ is separable, and let $P$ be the natural PVM on $\sA$ acting on $L^2(\sA)$. Let $\Hilbert_b$ be another separable Hilbert space and $E$ a POVM on the measurable space $\sZ$ acting on $\Hilbert_b$. Then the product POVM $P\otimes E$ exists, and if a quantum experiment with POVM $E$ and random outcome $Z$ is done on the conditional wave function $\psi'$, the joint distribution of $(A,Z)$ is $\scp{\psi}{P\otimes E|\psi}$.
\end{thm}

\begin{proof}
The distribution of $A$ is $\scp{\psi}{P\otimes I|\psi}$. Given $A$ and thus also $\psi'$, the distribution of $Z$ is $\scp{\psi'}{E|\psi'}$. Thus, a joint distribution exists and is given by, for any measurable $S\subseteq \sA \times \sZ$ and using the notation $S_a= \{z\in\sZ: (a,z) \in S\}$,
\begin{align}
\prob_\psi\Bigl( (A,Z)\in S \Bigr) 
&= \int_{\sA} da \: \|\psi(a)\|^2_b ~ \Bigl\langle \frac{\psi(a)}{\|\psi(a)\|_b} \Big|E(S_a)\Big| \frac{\psi(a)}{\|\psi(a)\|_b} \Bigr\rangle_b \\
&=\int_{\sA} da \, \scp{\psi(a)}{E(S_a)|\psi(a)}_b
\end{align}
(independently of the choice of representative of $\psi$). Since for fixed $S$, this is a bounded quadratic form in $\psi$, it is $\scp{\psi}{G(S)|\psi}$ for some operator $G(S)$, and $\sigma$-additivity of probabilities implies weak $\sigma$-additivity of $G(S)$, which is sufficient for a POVM. For $S=S'\times S''$, this reduces to $\scp{\psi}{P(S')\otimes E(S'')|\psi}$, so $G$ is the product POVM $P\otimes E$.
\end{proof}

\subsection{Proof of Corollary~\ref{cor:nS}}

We apply Theorem~\ref{thm:S} replacing $d\to Nd$ and $\R\to\R^N$ and obtain that the time evolution of $\psi$ in $\R^N$ exists for all $t>0$.

Let $J_N$ denote the $J$ operator of Section~\ref{sec:proofcorS}, defined now for the $Nd$-dimensional case; $J_N\psi_0$ is for arbitrary $\psi_0\in L^2(\R^N)$ a well-defined element of
\begin{align}
L^2 \bigl( [0,\infty)\times \partial(\R^N) \bigr) 
&\cong \oplus_i L^2 \bigl( [0,\infty)\times F_i \bigr)\\
&\cong L^2\bigl( [0,\infty)\times \bou  \times \{1,\ldots,N\} \times \R^{N-1} \bigr)\,,\label{piso}
\end{align}
where $\cong$ means isometrically isomorphic, and the last isomorphism is defined by the permutation $p$ of the variables as in \eqref{pdef}. A point in the ``configuration space'' is now $(T^1,\vX^1,I^1,x')$, and the joint density of $T^1,\vX^1,I^1$ is the appropriate marginal of $|J_N\psi_0|^2$ (ignoring the other $\vx_j$'s). Corollary~\ref{cor:nS} follows by repeating the same steps as in the proof of Corollary~\ref{cor:S}.

\subsection{Proof of Theorem~\ref{thm:collapse}}

Replacing $\{1,\ldots,N\}$ by $\sI$ with $N$ elements, we obtain operators $J_\sI$ and the permutation function $p:\cup_{i\in\sI} F_i \to \cup_{i\in\sI}\bou\times \R^{\sI\setminus\{i\}}$ now given by $p(x)=\bigl(\vx_i, (\vx_j: j\in\sI\setminus\{i\})\bigr)$ whenever $x\in F_i$. Further, we define the unitary isomorphism
\begin{align}
U_p:&~ L^2 \Bigl( [0,\infty)\times \partial(\R^\sI) \Bigr)
\to \bigoplus_{i\in\sI} L^2\Bigl( [0,\infty)\times \bou \times \R^{\sI\setminus\{i\}} \Bigr)\\
U_p(t,x)&= (t,p(x))
\end{align}
in analogy to \eqref{piso}.
The desired $\psi'$ of \eqref{psi'} then is the conditional wave function of $U_p J_\sI\psi_0$ as in Theorem~\ref{thm:cond} with Remark~\ref{rem:bundle} with $\sA=[0,\infty)\times\bou\times \sI$, $A=(T^1,\vX^1,I^1)$, $\Hilbert_b(i)=L^2(\R^{\sI\setminus\{i\}})$, and the $b$ variable corresponds to $x'=(\vx_j: j\in\sI\setminus\{i\})$. Given that $Z\neq \infty$, $\psi'$ is with probability 1 a well-defined element of $\Hilbert_b(i)$ with $i=I^1$.

\subsection{\RT{Proof of Theorem~\ref{thm:N}}}


\RT{Theorem~\ref{thm:collapse} and Corollary~\ref{cor:nS} show that $\psi^k$ and the distribution of $Z^k$ are almost surely well defined.} It remains to show that the distribution of $Z=(Z^1,\ldots,Z^N)$ comes from a POVM $E_\sI$. This follows from Theorem~\ref{thm:prod}, and $E_\sI$ can be specified recursively by setting, for $\sJ\subseteq \sI$,
\begin{align}
E_{\sJ}(\{(\infty\ldots\infty)\})
&= I-J_{\sJ}^* J_{\sJ}\\
E_{\sJ}(B)
&=J_{\sJ}^* U_p^* \Biggl( \bigoplus_{i\in\sJ} \Bigl[ P_{[0,\infty)\times\bou\times\{i\}} \otimes E_{\sJ\setminus\{i\}} \Bigr](B) \Biggr) U_p J_{\sJ}
\end{align}
for any measurable $B$ outside the sequences starting with $\infty$, that is, for $B\subseteq \bigl( [0,\infty)\times \bou \times \sI \bigr) \times \bigl( [0,\infty)\times \bou \times \sI \cup \{\infty\} \bigr)^{\#\sJ-1}$. Here, $P_\sA$ is the natural PVM on $\sA$. The end of the recursion is that, for every 1-element set $\sJ=\{j\}$,
\begin{align}
E_{\{j\}}(\{\infty\})&= I-J_{\{j\}}^* J_{\{j\}}\\
E_{\{j\}}(B) &= J_{\{j\}}^* \,  P_{[0,\infty)\times\bou}(B) \, J_{\{j\}}
\end{align}
for $B\subseteq [0,\infty)\times \bou$.

\bigskip

\noindent{\it Acknowledgments.} We thank Sascha Eichmann and Julian Schmidt for helpful discussions and two anonymous referees for pointing out errors in an earlier version of this article.

\section*{Declarations}

\noindent{\it Funding.} This work received no funding.

\noindent{\it Conflict of interests.} The authors declare no conflict of interest.

\noindent{\it Availability of data and material.} Not applicable.

\noindent{\it Code availability.} Not applicable.

\end{document}